\documentclass[10pt,conference,letterpaper]{IEEEtran}
\usepackage{epsfig,epsf,rotating,setspace,latexsym,amsmath,amssymb,amsfonts,bm,theorem,epstopdf,cite,authblk,bbm,mathrsfs,graphicx,caption}
\usepackage{subcaption}
\usepackage{algorithm}
\usepackage[noend]{algpseudocode}
\usepackage{color}
\usepackage{mathtools}
\usepackage{multirow}
\usepackage{soul}
\usepackage[letterpaper, left=0.625in, right=0.625in, bottom=1.03in, top=0.72in]{geometry}
\algrenewcommand\algorithmicforall{\textbf{foreach}}
\algrenewcommand\algorithmicindent{.8em}

\newtheorem{theorem}{Theorem}
\newtheorem{lemma}{Lemma}

\newtheorem{remark}{Remark}

\newtheorem{example}{Example}
\newenvironment{Proof}[1]{\medskip\par\noindent{\bf Proof:\,}\,#1}{{\mbox{\,$\blacksquare$}\par}}

\newcommand{\cq}{{\mathcal{Q}}}
\newcommand{\cR}{{\mathcal{R}}}
\newcommand{\cw}{{\mathcal{W}}}

\newcommand{\ck}{{\mathcal{K}}}

\newcommand{\cn}{{\mathcal{N}}}

\IEEEoverridecommandlockouts
\allowdisplaybreaks

\date{}
\title{Symmetric Private Information Retrieval (SPIR) on Graph-Based Replicated Systems}

\author{Shreya Meel \qquad Sennur Ulukus\\
\normalsize Department of Electrical and Computer Engineering\\
\normalsize University of Maryland, College Park, MD 20742 \\
\normalsize {\it smeel@umd.edu} \qquad {\it ulukus@umd.edu}}
 
\begin{document}

\maketitle

\begin{abstract}
We introduce the problem of symmetric private information retrieval (SPIR) on replicated databases modeled by a simple graph. In this model, each vertex corresponds to a server, and a message is replicated on two servers if and only if there is an edge between them. We consider the setting where the server-side common randomness necessary to accomplish SPIR is also replicated at the servers according to the graph, and we call this as message-specific common randomness. In this setting, we establish a lower bound on the SPIR capacity, i.e., the maximum download rate, for general graphs, by proposing an achievable SPIR scheme. Next, we prove that, for any SPIR scheme to be feasible, the minimum size of message-specific randomness should be equal to the size of a message. Finally, by providing matching upper bounds, we derive the exact SPIR capacity for the class of path and regular graphs. 
\end{abstract}

\section{Introduction}
In private information retrieval (PIR) \cite{chor}, a user wishes to download their desired message in a database replicated in multiple non-colluding servers without revealing the index of the message to any server. The notion of PIR capacity, i.e., the maximum ratio of the number of message and downloaded symbols, was introduced in \cite{rashmi_erasure_pir} and the exact PIR capacity for the original setting of \cite{chor} was found by Sun-Jafar \cite{SJ17}. This was followed by a line of work on PIR under various configurations (see \cite{ChaoTian, mdstpir,   banawan_pir_mdscoded, banawan_multimessage_pir, BU19, tian_leakage_pir, batuhan_hetero,  semantic_pir} and \cite{ulukusPIRLC} for a  survey). A drawback of PIR is that, it compromises the privacy of the messages that are not requested by the user. To deliver database privacy, symmetric PIR (SPIR) was formulated in \cite{spir_first}, which ensures that no information beyond the desired message is revealed to the user.

As shown in \cite{spir_first}, SPIR is not feasible unless some \emph{common randomness} is shared by the servers. The capacity of SPIR and the minimum amount of common randomness required for SPIR feasiblity, was characterized in \cite{c_spir} under the fully-replicated database setting.  Following this, several works studied SPIR under more practical settings; e.g., SPIR with MDS coded messages \cite{skoglund_mds_spir, 
 sun_spir_mds_mismatched}, SPIR with resilience against passive and active adversaries \cite{pir_spir_adversaries,  nan_eaves, nomeir_asymp_bspir}, SPIR for multiple messages \cite{wang_mmspir}, SPIR with side information\cite{zhusheng_spir_pir, chou_spir} and SPIR to retrieve a random message \cite{zhusheng_rspir}. So far, SPIR has been studied only in settings where each server stores all the messages in a coded or an uncoded form.

In this work, we propose an SPIR formulation on replicated databases, modeled by a graph as in the respective works on PIR \cite{graphbased_pir, SGT23, asymp_gxstpir , meel_multi_pir}. We focus on scenarios where fully replicating the databases is expensive, or the user has restricted access to the databases \cite{ali_dapac,  meel_hetdapac}. The graph-based replicated setting is a first step in this direction. In this model, each vertex corresponds to a server, and each edge represents a message stored on them. Further, we restrict the common randomness to be shared only by the servers sharing a message, one common randomness designated per message. This poses additional privacy constraint, since the randomness is now associated with the message through shared replication. Under this setup, we show that, the optimal (minimum) size of this randomness is equal to the length of a message. We propose an SPIR scheme that achieves the rate of $\frac{1}{N}$ for any graph with $N$ vertices. Further, we prove that our scheme is capacity-achieving for the class of $d$-regular (where $d$ denotes the degree of each vertex) and path graphs, by deriving matching upper bounds. For these classes of graphs, we find that the additional constraint of database privacy does not hurt the PIR capacity by more than half.

\section{System Model}
We consider a database of $K\geq2$ independent messages $\mathcal{W}=\{W_1,\ldots,W_K\}$, each comprising $L$ independent symbols chosen uniformly at random from a finite field $\mathbb{F}_q$, 
\begin{align}
    H(\cw)&=H(W_1)+\ldots+H(W_K) \\
        &=KL ,\ \text{ in } q \text{-ary units.}
\end{align}
The messages are stored on $N\geq 2$ non-colluding servers. Each message $W_k\in \cw$ is replicated exactly twice and stored on two distinct servers in $[N]$. Such a 2-replicated message system can be represented by a simple graph $G=(V,E)$ where each vertex in $V$ represents a server and each edge in $E$ represents a message. An edge is associated with two vertices if and only if a message is replicated on the two corresponding servers. In this work, we assume that $G$ is connected, i.e., there exists a path between each pair of vertices.

\begin{figure}[t]
\centering
    \begin{subfigure}{0.5\textwidth}
    \centering
    \includegraphics[width=0.8\textwidth]{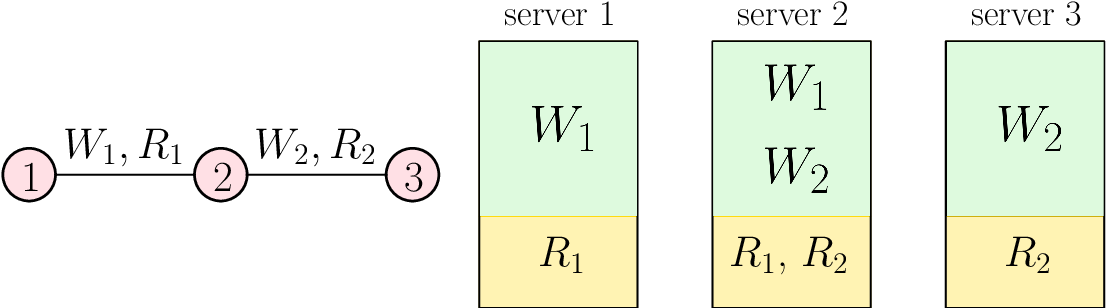} 
    \subcaption{$\mathbb{P}_3$}
    \vspace{2mm}
    \label{fig:spir_p3}
    \end{subfigure}
    \begin{subfigure}{0.5\textwidth}
    \centering
    \includegraphics[width=0.8\textwidth]{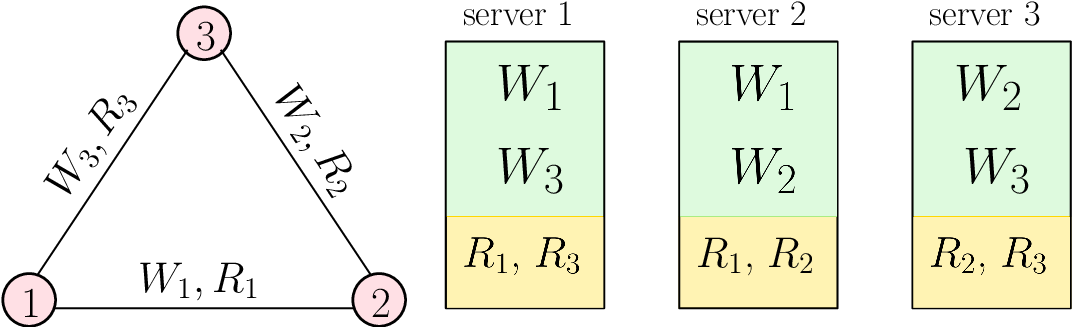}
    \subcaption{$\mathbb{C}_3$}
    \vspace{2mm}
    \label{fig:spir_c3}
    \end{subfigure}
    \caption{SPIR system model for path $\mathbb{P}_3$, and cyclic $\mathbb{C}_3$ graphs.}
    \label{fig:sys_mod}
\end{figure}

In this work, for each $k\in [K]$, we endow the servers sharing $W_k$ with a private random variable $R_k$, independent of $W_k$, and whose realization is unavailable to the user and to the servers that do not store $W_k$. This can arise when sharing randomness variables across all servers is restricted due to communication constraints. We refer to $R_k$ as the \emph{message-specific common randomness}. Clearly, $\mathcal{R} =\{R_1,\ldots,R_K\}$ is replicated according to the same $G$, and each $R_k$ is also $2$-replicated. Moreover, we assume that $R_k,~k\in [K]$ are independent and identically distributed (i.i.d.). Fig.~\ref{fig:sys_mod} illustrates the storage across servers for the SPIR systems corresponding to the simple path $\mathbb{P}_N$ and cyclic $\mathbb{C}_N$ graphs.

Let $\theta$ represent the desired message index and $\cq$ represent the private randomness in the schemes followed by the user to retrieve the $K$ messages in the system. Since $\cq$ is decided prior to choosing the message index, it is independent of $\theta$. Further, $\theta$ and $\cq$ are independent of $\mathcal{W}$ and $\mathcal{R}$, since the user has no information of the content stored at the servers. 

Suppose $\theta=k$. To retrieve $W_k$, the user privately generates $N$ queries $Q_1^{[k]}, \ldots, Q_N^{[k]}$ using $\cq$, i.e.,
\begin{align}\label{eq:query deterministic of query randomness}
    H(Q_1^{[k]},\ldots,Q_N^{[k]}|\cq) = 0,
\end{align}
and sends $Q_n^{[k]}$ to server $n$. Upon receiving the query, server $n$ responds with an answer $A_n^{[k]}$. Let $\mathcal{W}_n$ and $\mathcal{R}_n$ denote the set of messages and randomness stored at server $n$. Then, $A_n^{[k]}$ is a deterministic function of $Q_n^{[k]}$, $\mathcal{W}_n$ and $\mathcal{R}_n$, i.e.,
\begin{align}\label{eq:answer of server deterministic}
    H(A_n^{[k]}|Q_n^{[k]}, \mathcal{W}_n, \mathcal{R}_n) = 0.
\end{align}
Next, we state the requirements of our SPIR problem: user privacy, reliability and database privacy. For user privacy, the query and answer for each server are identically distributed, irrespective of $\theta$, i.e., for every $n\in [N]$ and any index $k$,
\begin{align}\label{eq:user_privacy}
(Q_n^{[k]}, A_n^{[k]},\mathcal{W}_n, \mathcal{R}_n) \sim (Q_n^{[1]}, A_n^{[1]},\mathcal{W}_n, \mathcal{R}_n).
\end{align}
To guarantee reliability, the user should be able to exactly recover their requested message $W_{k}$, using the answers from all the servers, i.e.,
\begin{align}
    H(W_k|A_1^{[k]},\ldots,A_N^{[k]}, \cq) = 0.\label{eq:reliability}
\end{align}
Finally, to ensure database privacy, we require that, even if the knowledge of common randomness designated for a subset of messages is available, the answers reveal no information on the subset of undesired messages, if the common randomness corresponding to them is unavailable. Thus, for any subset $\mathcal{J}\subseteq [K]\setminus \{k\}$, the answers and queries should satisfy, 
\begin{align}
    I( W_{\mathcal{J}};A_{1:N}^{[k]},Q^{[k]}_{1:N},\cR\setminus R_{\mathcal{J}},\cw\setminus \{W_k,W_{\mathcal{J}}\},\cq)=0,\label{eq:database_privacy}
\end{align}
where $W_{\mathcal{J}}=\{W_\ell:\ell\in \mathcal{J}\}$, $R_{\mathcal{J}}=\{R_\ell:\ell\in \mathcal{J}\}$, $Q_{1:N}^{[k]}=\{Q_1^{[k]},\ldots,Q_N^{[k]}\}$ and $A_{1:N}^{[k]}=\{A_1^{[k]},\ldots,A_N^{[k]}\}$.

\begin{remark}
If in \eqref{eq:database_privacy}, we let $\mathcal{J}=[K]\setminus \{k\}$, we obtain $
I({W}_{\overline{k}};A^{[k]}_{1},\ldots,A_N^{[k]},Q^{[k]}_1,\ldots,Q_N^{[k]},R_k,\cq)=0$ which means that the answers from the servers given the queries, reveal no information on ${W}_{\overline{k}}:=\mathcal{W}\setminus W_k$ to the user, even if $R_k$ is retrieved in the process, which differs from the definition in \cite{c_spir}, due to integration of message-specific common randomness.
\end{remark}

An SPIR scheme is said to be achievable if it simultaneously satisfies \eqref{eq:user_privacy}, \eqref{eq:reliability}, and \eqref{eq:database_privacy}. The following two metrics quantify the efficiency of an SPIR scheme.

\paragraph*{Capacity $\mathscr{C}(G)$} 
The rate of an SPIR scheme $T$ on $G$ is the ratio of the number of desired message symbols and the total number of downloaded symbols. The SPIR capacity of $G$ is defined as,
\begin{align}
    \mathscr{C}\left(G\right) \triangleq \sup_{T} \frac{L}{\sum_{n=1}^N H(A_n^{[k]})},
\end{align}
where the supremum is over all possible schemes $T$ on $G$.

\paragraph*{Randomness ratio $\rho$} In our model, the common randomness is message-specific and is stored on servers according to the graph $G$ (see Fig.~\ref{fig:sys_mod}). This is different from the original SPIR formulation which assumes that the common randomness is available to all servers. To account for this, we define the \emph{randomness ratio},
\begin{align}
    \rho \triangleq \frac{H(R_k)}{L}, \quad k\in [K]
\end{align} 
as the size of message-specific randomness relative to the size of a message, required for an achievable SPIR scheme.  

\section{Main Results}
In this section, we present our main results. Note that, if $N=2$, the only connected simple graph is $\mathbb{P}_2$. That is, a single message is replicated on two servers. The PIR and SPIR problems become trivial and the capacity is $1$ in both cases. We hereby focus on graphs with $N\geq 3$. 
\begin{theorem}\label{thm:achievable_rate_all_G}
    For any graph $G$ with $N$ vertices, its SPIR capacity $\mathscr{C}(G)$ can be bounded as
    \begin{align}
        \mathscr{C}(G)\geq \frac{1}{N},
    \end{align}
provided that the randomness ratio $\rho=1$.
\end{theorem}
The proof of Theorem \ref{thm:achievable_rate_all_G} follows from a scheme construction with rate $\frac{1}{N}$, which is presented in Section \ref{sec:scheme1}. 
\begin{remark}
Similar to an achievable PIR rate on graph-replicated databases \cite{graphbased_pir,  SGT23,  BU19, our_journal2025}, the SPIR rate achieved by Theorem \ref{thm:achievable_rate_all_G} on $G$ is strictly decreasing in $N$.
\end{remark}
\begin{theorem}\label{thm:bound_rho}
For any SPIR scheme, the required randomness ratio $\rho$ is at least $1$; otherwise SPIR is not feasible.
\end{theorem}
\begin{theorem}\label{thm:upperbnd_capacity}
    If $G=\mathbb{P}_N$ or $G$ is a $d$-regular graph, 
    \begin{align}
        \mathscr{C}(G)\leq \frac {1}{N}.
    \end{align}
\end{theorem}
Theorems \ref{thm:achievable_rate_all_G} and \ref{thm:upperbnd_capacity} imply that $\mathscr{C}(G)=\frac{1}{N}$ for these graphs. The proofs of Theorems \ref{thm:bound_rho} and \ref{thm:upperbnd_capacity} appear in Section \ref{sec:proofs}. 

\begin{remark}
    The PIR capacity for $\mathbb{P}_N$ is $\frac{2}{N}$ \cite{our_journal2025}. Therefore, incorporating the database privacy constraint \eqref{eq:database_privacy} hurts the capacity by exactly half.
\end{remark}
\begin{remark}
    The PIR capacity for $\mathbb{C}_N$ is $\frac{2}{N+1}$ \cite{BU19}. Since $\mathbb{C}_N$ is a $2$-regular graph, the corresponding SPIR capacity is $\frac{1}{N}$, which is greater than half of its PIR capacity.
\end{remark}
\begin{remark}
In general, the PIR capacity for regular graphs with $N$ vertices is bounded above by $\frac{2}{N}$ \cite{SGT23}. The corresponding SPIR capacity therefore, is at least half the PIR capacity.
\end{remark}
\begin{remark}
Regular graphs with equal $N$ and varying $K$ have equal SPIR capacities, which is not necessarily true for PIR. For instance, the cyclic graph $\mathbb{C}_N (d=2)$ with $K=N$ has PIR capacity $\frac{2}{N+1}$, while for the complete graph $\mathbb{K}_N (d=N-1)$ with $K=\binom{N}{2}$, no scheme is known to achieve this bound for $N\geq 4$, and the PIR capacity, in general, is open.
\end{remark}

\begin{remark}
If we replace every edge of a graph $G$ with $r$ parallel edges, the resulting graph structure is an $r$-multigraph, denoted by $G^{(r)}$. The problem of PIR on multigraph-based replicated systems was recently explored in \cite{meel_multi_pir}, and the exact capacity of $r$-multipath $\mathbb{P}_N^{(r)}$ was derived to be  $\frac{1}{N(1-2^{-r})}$ for even $N$. Interestingly, when $r\to \infty$, this quantity matches the SPIR capacity for $\mathbb{P}_N$. This indicates that, if the number of messages shared between two servers $r$ is arbitrarily large, the information that the user learns from any PIR scheme about the undesired messages becomes arbitrarily small.
\end{remark}
\section{Proof of Theorem \ref{thm:achievable_rate_all_G}}\label{sec:scheme1}
The scheme achieving the rate $\frac{1}{N}$ with $\rho=1$ is inspired from Raviv et al.'s PIR scheme \cite{graphbased_pir} on 2-replicated systems, coupled with one-time padding \cite{shannon_otp}, for database privacy.

Let $I(G)$ denote the incidence matrix of the graph $G$. That is, given $G$, $I(G)$ is defined as the $|V|\times |E| = N\times K$ binary matrix, where rows represent vertices and columns represent the edges. The $(n,\ell)$-th entry of $I(G)$ is 1 if edge $\ell$ is incident with vertex $n$ and $0$ otherwise.

For the SPIR system based on $G$, suppose each message and randomness is a single symbol of $\mathbb{F}_q$, where $R_\ell$ is picked uniformly at random from $\mathbb{F}_q$ for all $\ell\in [K]$. For each server $n\in [N]$, we denote the message indices it holds, in ascending order, by the ordered set $\mathcal{F}_n = (\ell:W_\ell\in \mathcal{W}_n)$. Further, we represent $\mathcal{W}_n$ as a vector $\bm{W}_n = [W_\ell, \ell\in \mathcal{F}_n]^{\top}$.

Each column of $I(G)$ has exactly two $1$'s. Let us write the signed incidence matrix $\bar{I}(G)$ by replacing the lower $1$-entry to $-1$ for each column. Suppose $\theta=k$. To privately retrieve $W_k$, the user chooses $K$ symbols $(h_1, h_2, \ldots, h_K)$ independently and uniformly at random from $\mathbb{F}_q$, and forms the matrix:
 \begin{align}\label{eq:query_matrix}
     \bm{H} = \bar{I}(G)\cdot \text{diag}(h_1,\ldots,h_K).
 \end{align}
Let $\bm{h}_n$ denote the $n$-th row of $\bm{H}$ after discarding the zeros. Then, the user sends the following queries, where $\bm{e}_m$ is the standard unit column vector with $1$ at the $m$-th coordinate:
\begin{align}\label{eq:queries_scheme1}
    Q_n^{[k]} = \begin{cases}
        \bm{h}_n^{\top}, & n \in [N]\setminus \{j\}\\
        \bm{h}_n^{\top} + \bm{e}_m, & n = j,
    \end{cases}
\end{align}
 assuming that $W_k$ is replicated at servers $i$ and $j$ and that $\bm{e}_m^{\top}\bm{W}_j=W_k$. 
Server $n$, upon receiving $Q_n^{[k]}$ responds with the following answer:
\begin{align}
    A_n^{[k]} = Q_n^{[k]\top}\bm{W}_n + \sum_{\ell\in \mathcal{F}_n}\bar{I}(G)(n,\ell)\cdot R_\ell.
\end{align}
Now, to recover $W_k$, the user computes the sum of the answers. By the design of queries and answers, 
\begin{align}\label{eq:scheme_answers}
A_n^{[k]} = \begin{cases}
    \sum_{\ell\in \mathcal{F}_n} \bar{I}(G)(n,\ell) (h_\ell W_\ell + R_{\ell}) ,& \!\!\!\! \!\! n\in [N]\setminus \{j\}\\
    \sum_{\ell\in \mathcal{F}_n} \bar{I}(G)(n,\ell) (h_\ell W_\ell + R_{\ell})+W_k, &n = j.   
\end{cases}
\end{align}
Summing the answers from servers $n\in [N]$ gives,
\begin{align}
  \sum_{n\in [N]}&\left(\sum_{\ell\in \mathcal{F}_n} \bar{I}(G)(n,\ell) (h_\ell W_\ell + R_{\ell})\right)+W_k \nonumber\\
    &= \sum_{\ell\in \mathcal{F}_n} (h_\ell W_\ell + R_{\ell})\left( \sum_{n\in [N]} \bar{I}(G) (n,\ell) \right) + W_k \\
    &=W_k,  \label{eq:row sum of I bar G is 0}
\end{align}
where \eqref{eq:row sum of I bar G is 0} is because the entries of any column $\ell$ of $\bar{I}(G)$ sum to 0. This proves reliability \eqref{eq:reliability}.

\begin{remark}
The answer generation of our scheme bears some similarity with the scheme of secure summation eg., \cite{zhao_securesum}. Since we compute the sum of answers for decoding the desired message, the idea of utilizing randomness symbols which sum across the servers to zero, is a common thread in both the schemes.
\end{remark}

For any desired message index $k$, server $n$ receives a query vector of length $\delta(n)$, where $\delta(n)$ is the degree of vertex $n$ in $G$. Thus, the server observes a uniformly distributed random vector over $\mathbb{F}_q^{\delta(n)}$. To compute the answer, server $n$ combines its stored messages with the query coefficients and adds a linear combination of the stored randomness symbols, with coefficients from $\bar{I}(G)$. Hence, the user privacy constraint \eqref{eq:user_privacy} is satisfied.

To see that \eqref{eq:database_privacy} holds, besides $W_k$, the user receives linear combinations of $W_\ell,$ combined with $R_\ell$ (with suitable signs), $ \ell\in [K]\setminus \{k\}$ and $R_k$. Then, by the i.i.d. nature of $R_\ell$, independence of $\cR$ from $\cq$ and $\cw$
and the one-time pad theorem \cite{shannon_otp}, the user learns no information on any message subset $W_{\mathcal{J}}$ beyond $W_k$. 

\paragraph*{Rate} From each of the $N$ servers, the user downloads a single symbol of $\mathbb{F}_q$ as answer, to recover $L=1$ symbol of the desired message. This results in an SPIR scheme for $G$ with rate $\frac{1}{N}$. Further, $\rho=1$ since $H(R_k)=L, \forall k\in [K]$.

Next, we illustrate the scheme for some families of graphs.

\begin{example}
Consider the path graph $\mathbb{P}_3$, as shown in Fig.~\ref{fig:spir_p3} where each message and randomness consists of a single symbol from $\mathbb{F}_q$. The matrices $I(\mathbb{P}_3)$ and $\bar{I}(\mathbb{P}_3)$ are:
\begin{align}
    I(\mathbb{P}_3)=
    \begin{bmatrix}
        1 & 0 \\
        1 & 1\\
        0 & 1
    \end{bmatrix},\quad
        \bar{I}(\mathbb{P}_3)=
    \begin{bmatrix}
        1 & 0 \\
        -1 & 1\\
        0 & -1
    \end{bmatrix}.
\end{align}The user chooses two random symbols $h_1$ and $h_2$, and forms the matrix
\begin{align}
    \bm {H}=
    \begin{bmatrix}
        h_1 & 0\\
        -h_1 & h_2\\
        0 & -h_2
    \end{bmatrix}.
\end{align}
Then, the queries sent are as follows:
\begin{align}
    Q_1^{[\theta]}=h_1, \ Q_2^{[\theta]} = [-h_1, \quad h_2]^\top+\bm{e}_{\theta}, \ Q_3^{[\theta]} =-h_2.
\end{align}
The answers returned by the servers are:
\begin{align}
    A_1^{[\theta]}&=h_1W_1+R_1,\notag\\
    A_2^{[\theta]}&= -h_1W_1+h_2W_2+W_{\theta}-R_1+R_2,\notag\\
    A_3^{[\theta]}&= -h_2W_2-R_2.
\end{align}To decode $W_{\theta}$, the user computes the sum of all the answers, resulting in the rate $\frac{1}{3}$.
\end{example}

\begin{example}
Consider the cyclic graph $\mathbb{C}_3$ as shown in Fig.~\ref{fig:spir_c3}, with $L=1$ and $\rho=1$. The matrices $I(\mathbb{C}_3)$ and $\bar{I}(\mathbb{C}_3)$ are:
\begin{align}
    I(\mathbb{C}_3)=
    \begin{bmatrix}
        1 & 0 & 1\\
        1 & 1 & 0\\
        0 & 1 & 1
    \end{bmatrix},\quad
        \bar{I}(\mathbb{C}_3)=
    \begin{bmatrix}
        1 & 0 & 1\\
        -1 & 1 & 0\\
        0 & -1 & -1
    \end{bmatrix}.
\end{align}
 The user chooses the random symbols $h_1,h_2,h_3$ from $\mathbb{F}_q$ and sends queries according to \eqref{eq:queries_scheme1}. The answers returned by the servers when $\theta=1$ are:
\begin{align}
    A_1^{[1]}&= h_1W_1+h_3W_3+R_1+R_3, \notag\\
    A_2^{[1]}&=-h_1W_1+h_2W_2+W_1-R_1+R_2, \notag\\
     A_3^{[1]}&=-h_2W_2-h_3W_3-R_2-R_3.
\end{align}
Summing the answers, the user decodes $W_1$.
\end{example}
The same SPIR rate of $\frac{1}{3}$ is achieved for both $\mathbb{P}_3$ and $\mathbb{C}_3$. By Theorem \ref{thm:upperbnd_capacity}, our scheme on path and cyclic graph, is capacity-achieving.  

\begin{figure}[t!]
    \centering
    \begin{subfigure}[t!]{0.49\textwidth}
    \centering
    \includegraphics[width=0.36\linewidth]{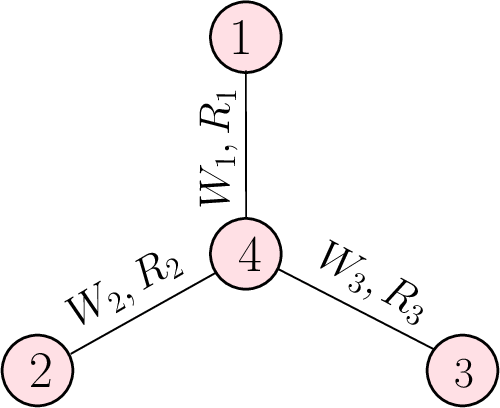}
    \subcaption{Star graph $\mathbb{S}_4$}
    \vspace{2mm}
    \label{fig:spir_s4}
    \end{subfigure}%
    \hfill
    \begin{subfigure}[t!]{0.49\textwidth}
    \centering
    \includegraphics[width=0.5\linewidth]{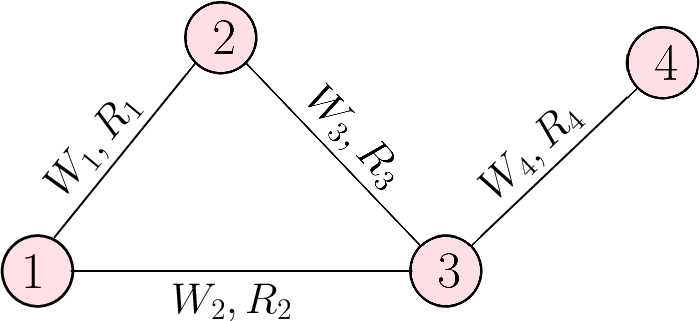}
    \subcaption{Graph $\mathbb{M}$}
    \label{fig:spir_g4}
    \end{subfigure}
    \caption{SPIR systems with $N=4$ servers.}
\end{figure}

\begin{example}
 Consider the star graph $\mathbb{S}_4$ as shown in Fig. ~\ref{fig:spir_s4}, with $L=1$ and $\rho=1$. The matrices $I(\mathbb{S}_4)$ and $\bar{I}(\mathbb{S}_4)$ are:
\begin{align}
    I(\mathbb{S}_4)=
 \begin{bmatrix}
        1 & 0 & 0\\
        0 & 1 & 0\\
        0 & 0 & 1\\
        1 & 1 & 1
    \end{bmatrix},\quad
        \bar{I}(\mathbb{S}_4)=
    \begin{bmatrix}
        1 & 0 & 0\\
        0 & 1 & 0\\
        0 & 0 & 1\\
        -1 & -1 & -1
    \end{bmatrix}.    
\end{align}To retrieve the desired message $W_{\theta}$, the answers returned are: $h_1W_1+R_1, h_2W_2+R_2, h_3W_3+R_3, -(h_1W_1+h_2W_2+h_3W_3+R_1+R_2+R_3)+W_{\theta}$
by servers $1,2,3$ and $4$ respectively. Clearly, the rate is $\frac{1}{4}$.
\end{example}
\begin{example}

    Consider the SPIR system on the graph $\mathbb{M}$ in Fig.~\ref{fig:spir_g4}, with $L=1$ and $\rho=1$. Its signed incidence matrix is, 
    \begin{align}
    \bar{I}(\mathbb{M})=
    \begin{bmatrix}
        1 & 1 & 0 & 0\\
        -1 & 0 & 1 & 0 \\
        0 & -1 & -1 & 1\\
        0 & 0 & 0 & -1   
    \end{bmatrix}.
    \end{align}
The user chooses the random symbols $h_1$,$h_2$, $h_3$ and $h_4$, from $\mathbb{F}_q$ and accordingly sends the queries to the servers using \eqref{eq:queries_scheme1}. For example, if $\theta=3$, the answers returned are:
\begin{align}
    A_1^{[3]} &= h_1W_1+h_2W_2+R_1+R_2\notag\\
    A_2^{[3]} &= -h_1W_1+h_3W_3-R_1+R_3\notag\\
    A_3^{[3]} &= -h_2W_2-h_3W_3+W_3+h_4W_4-R_2-R_3+R_4 \notag\\
    A_4^{[3]} &= -h_4W_4-R_4.
\end{align}
The user can decode $W_3$ by computing the sum of the answers, and the resulting rate is $\frac{1}{4}$.
\end{example}
Both $\mathbb{S}_4$ and $\mathbb{M}$ achieve the same SPIR rate, despite their different structures.
\section{Proofs of Theorems \ref{thm:bound_rho} and \ref{thm:upperbnd_capacity}}\label{sec:proofs}

We make the following observation. Since the scheme is known globally, the user can perform the answer generation of server $n$ and obtain $R_j\in \cR_n$ from the received answer, given $\cw_n$ and $\cR_n\setminus \{R_j\}$, i.e.,
\begin{align}
   H(R_j|A_n^{[k]}, \cw_n,\cR_n\setminus\{R_j\},Q_n^{[k]})=0.\label{eq:R_k_deterministic} 
\end{align} Now, we state the following lemmas. The first lemma is an extension of \cite[Lemma 1]{c_spir} and \cite[Proposition 2]{SGT23} to our setting. 
\begin{lemma}\label{lem:answers_independent_of_index}
    For any subsets $\mathcal{J},\ck\subseteq [K]$, let $W_{\mathcal{J}}=\{W_{\ell}:\ell\in \mathcal{J}\}$ and $R_{\ck}=\{R_{\ell}:\ell\in \ck\}$. Then, for any server $n\in [N]$ and any $k,k'\in [K]$,
    \begin{align}
        H(A_n^{[k]}|W_{\mathcal{J}},R_{\ck}, Q_n^{[k]})=H(A_n^{[k']}|W_{\mathcal{J}},R_{\ck}, Q_n^{[k']}).
    \end{align}
\end{lemma}
\begin{Proof}
    The proof follows from the user privacy constraint \eqref{eq:user_privacy} of server $n$ and the fact that $A_n^{[k]}$  does not depend on the part of $W_{\mathcal{J}}$ and $R_{\mathcal{K}}$ not intersecting $\mathcal{W}_n$ and $\mathcal{R}_n$, respectively.
\end{Proof}

The next lemma is an extension of \cite[Lemma 2]{c_spir}.
\begin{lemma}\label{lem:answer_indep_randomness_given_query}
  For any subsets $\mathcal{J},\ck\subseteq [K]$, 
  \begin{align}
      H(A_n^{[k]}|W_{\mathcal{J}},R_{\ck}, Q_n^{[k]})=H(A_n^{[k]}|W_{\mathcal{J}},R_{\ck}, Q_n^{[k]},\cq).
  \end{align}
\end{lemma}
\begin{Proof}
    The proof involves showing that 
    \begin{align}
        I(A_n^{[k]},\cw_n^c,\cR_n^c;\cq|W_{\mathcal{J}},R_{\ck}, Q_n^{[k]})=0,
    \end{align} where  $\cw_n^{c}:=\cw_n\setminus W_{\mathcal{J}}$ and $\cR_n^{c}:=\cR_n\setminus R_{\mathcal{K}}$.
\end{Proof}

The next lemma is an extension of \cite[Lemma 3]{SGT23} modified to accommodate database privacy \eqref{eq:database_privacy}.
\begin{lemma}
\label{lem:sum_entropy of answers}
 For any two servers $i$ and $j$ that share the message $W_k$ and randomness $R_k$, the following holds: 
    \begin{align}
        H(A_i^{[k]}|&\cq)+H(A_j^{[k]}|\cq)\notag\\
        \geq &H(A_i^{[k]}|R_{\overline{k}},W_{\overline{k}},\cq)+H(A_j^{[k]}|R_{\overline{k}},W_{\overline{k}},\cq)\\
        \geq &(1+\rho)L,
    \end{align} 
    where $R_{\overline{k}}:=\cR\setminus \{R_k\}$.
\end{lemma}
\begin{Proof} We have that $H(W_k,R_k)=(1+\rho)L$. Next, conditioning on $W_{\overline{k}},R_{\overline{k}}$ and $\cq$, using \eqref{eq:reliability} and the fact that $H(R_k|A_i^{[k]},A_j^{[k]},R_{\overline{k}},\cw,\cq)=0$, completes the proof.
\end{Proof}

Now, we proceed with the proof of Theorem \ref{thm:bound_rho}.

\subsection{Proof of Theorem \ref{thm:bound_rho}} \label{subsec:proof_thm2}
Let $W_k,R_k$ be stored on servers $i$ and $j$. For the desired message index, $k'\neq k$, from database privacy \eqref{eq:database_privacy}, choosing $\mathcal{J}=
    \{k\}$, we have
\begin{align}
0=& I(W_k;A_1^{[k']},\ldots,A_N^{[k']},R_{\overline{k}},W_{\overline{k}}\setminus \{W_{k'}\},\cq)\\
=&  I(W_k;A_1^{[k']},\ldots,A_N^{[k']}|R_{\overline{k}},W_{\overline{k}}\setminus \{W_{k'}\},\cq)\notag\\
&+I(W_k;W_{k'}|A_1^{[k']},\ldots,A_N^{[k']},R_{\overline{k}},W_{\overline{k}}\setminus \{W_{k'}\},\cq)\label{eq:decodability of W_k'}\\
=& I(W_k;W_{k'},A_1^{[k']},\ldots,A_N^{[k']}|R_{\overline{k}},W_{\overline{k}}\setminus \{W_{k'}\},\cq)\\
=& I(W_k;A_1^{[k']},\ldots,A_N^{[k']}|R_{\overline{k}},W_{\overline{k}},\cq)\label{eq:independence_of_messages}\\
=& I(W_k;A_i^{[k']},A_j^{[k']}|R_{\overline{k}},W_{\overline{k}},\cq)\label{eq:answers other than i and j exactly deterministic}\\
\geq & I(W_k;A_i^{[k']}|R_{\overline{k}},W_{\overline{k}},\cq)\\
=& I(W_k;A_i^{[k]}|R_{\overline{k}},W_{\overline{k}},Q_i^{[k]})\label{eq:consequence of lemma 1}\\
=&H(A_i^{[k]}|W_{\overline{k}},R_{\overline{k}},Q_i^{[k]})-H(A_i^{[k]}|\cw,\cR,Q_i^{[k]})\notag\\
&-I(R_k;A_i^{[k]}|\cw,R_{\overline{k}},Q_i^{[k]})\\
=&H(A_i^{[k]}|W_{\overline{k}},R_{\overline{k}},Q_i^{[k]})-I(R_k;A_i^{[k]}|\cw,R_{\overline{k}},Q_i^{[k]})\label{eq:Ai_deterministic given everything else}\\
=& H(A_i^{[k]}|W_{\overline{k}},R_{\overline{k}},Q_i^{[k]})-H(R_k)\label{eq:R_k_deterministic given Ai}\\
= & H(A_i^{[k]}|W_{\overline{k}},R_{\overline{k}},\cq)-H(R_k), \label{eq:answer i_randomness}
\end{align}
where \eqref{eq:decodability of W_k'} follows since $I(W_k;W_{k'}|A_1^{[k']},\ldots,A_N^{[k']},R_{\overline{k}},W_{\overline{k}}\setminus \{W_{k'}\},\cq)=0$ by \eqref{eq:reliability}, \eqref{eq:independence_of_messages} follows by the independence of messages, \eqref{eq:consequence of lemma 1} is a consequence of \eqref{eq:query deterministic of query randomness}, Lemma \ref{lem:answers_independent_of_index} and Lemma \ref{lem:answer_indep_randomness_given_query}, \eqref{eq:Ai_deterministic given everything else} is by \eqref{eq:answer of server deterministic}, \eqref{eq:R_k_deterministic given Ai} is due to \eqref{eq:R_k_deterministic} and \eqref{eq:answer i_randomness} is due to Lemma \ref{lem:answer_indep_randomness_given_query} and \eqref{eq:query deterministic of query randomness}. 
Similarly, we have
\begin{align}\label{eq:answer j_randomness}
H(A_j^{[k]}|W_{\overline{k}},R_{\overline{k}},\cq)-H(R_k)\leq0.
\end{align}
Adding \eqref{eq:answer i_randomness} and \eqref{eq:answer j_randomness}, we get $2H(R_k)\geq H(A_i^{[k]}|W_{\overline{k}},R_{\overline{k}},\cq)+H(A_j^{[k]}|W_{\overline{k}},R_{\overline{k}},\cq)\geq (1+\rho)L$
by Lemma \ref{lem:sum_entropy of answers}. Substituting $H(R_k)=\rho L$, we obtain $\rho\geq 1$.

\subsection{Proof of Theorem \ref{thm:upperbnd_capacity}}\label{subsec:proof_thm3}
\textbf{$d$-Regular graph $G$:} Note that, to respect user privacy \eqref{eq:user_privacy}, the result of Lemma \ref{lem:sum_entropy of answers}, combined with Theorem \ref{thm:bound_rho},
\begin{align}\label{eq:pair_ans_bound}
    H(A_i^{[k]}|\cq)+H(A_j^{[k]}|\cq)\geq 2L
\end{align} 
should hold for every pair of servers $(i,j)$ which share a file, irrespective of the desired index $k$. Summing \eqref{eq:pair_ans_bound} over all $i,j$,
\begin{align}
    \sum_{(i,j)\in E}  H(A_i^{[k]}|\cq)+H(A_j^{[k]}|\cq)&\geq 2KL 
    \end{align}
    which, because $G$ is $d$-regular yields
\begin{align}
    & d\left(\sum_{n=1}^N H(A_n^{[k]}|\cq)\right) \geq 2KL.
\end{align}
Then, by $Nd=2K$, this results in 
\begin{align}
    \frac{L}{\sum_{n=1}^N H(A_n^{[k]})}\leq  \frac{L}{\sum_{n=1}^N H(A_n^{[k]}|\cq)}\leq\frac{d}{2K} = \frac{1}{N}.
\end{align}

\textbf{Path graph $\mathbb{P}_N$:}
To show the upper bound for paths, we need the SPIR version of \cite[Theorem 6]{SGT23}, as given by the following lemma. It bounds the answer size from a server with respect to the answers from servers in its neighbor set, i.e., the servers with which it shares a message and randomness.
\begin{lemma}
\label{lem:single_server_answer_entropy}
 For a server $S\in [N]$, with degree $\delta$ in $G$, let $\cn(S)=\{S_1,\ldots,S_{\delta}\}$ denote its neighbor set. Then, for any $k\in [K]$,
\begin{align}\label{eq:lem_ineq}
    H(A_S^{[k]}|\cq)\geq \sum_{i=1}^{\delta}\max\left\{0,2L-\sum_{j=i}^\delta H(A_{S_j}^{[k]}|\cq)\right\}.
\end{align}
\end{lemma}
\begin{Proof} 
In this proof, for every $i\in [\delta]$, let $W_{i}$ and $R_{i}$, respectively denote the message and randomness stored on servers $S$ and $S_i\in \cn(S)$. Let $\cw^{c}:=\cw\setminus \{\cup_{i=1}^{\delta}{W}_i\}$ and $\cR^c:=\cR\setminus \{\cup_{i=1}^{\delta}R_i\}$. Conditioning on $\cw^c$ and $\cR^c$, we obtain
\begin{align}
 H(&A_S^{[k]}|\cq)\notag\\
 =&I(A_S^{[k]};W_{[\delta]},R_{[\delta]}|\cq,\cw^c,\cR^c)\label{eq:follows from A_s deterministic}\\
\geq & \sum_{i=1}^{\delta}\left(2L-H(W_i,R_i|A_S^{[i]},W_{[i-1]},R_{[i-1]},\cq,\cw^c,\cR^c)\right)\label{eq:follows by Thm2},
\end{align}
where \eqref{eq:follows from A_s deterministic} is due to \eqref{eq:answer of server deterministic} and \eqref{eq:follows by Thm2} follows from Lemmas \ref{lem:answers_independent_of_index}, \ref{lem:answer_indep_randomness_given_query}, and Theorem \ref{thm:bound_rho}. 
Next, using \eqref{eq:R_k_deterministic}, we upper bound the second term for each $i$ in the sum of \eqref{eq:follows by Thm2} by $\sum_{j=i}^{\delta}H(A_{S_j}^{[k]}|\cq)$ and obtain \eqref{eq:lem_ineq} from the non-negativity of mutual information.
\end{Proof}

To show the SPIR capacity upper bound $\mathscr{C}(\mathbb{P}_N)\leq\frac{1}{N}$, we consider the cases of $N$ even and odd separately. If $N$ is even, let $g$ be a positive integer such that $N=2g$, hence
\begin{align}\label{eq:applying_adjacency_bound}
    \sum_{n=1}^N H(A_n^{[k]})=&\sum_{j=1}^{g} H(A^{[k]}_{2j-1})+H(A^{[k]}_{2j})
    \geq \sum_{j=1}^g 2L 
\end{align}
where \eqref{eq:applying_adjacency_bound} follows from \eqref{eq:pair_ans_bound} since servers $(2j-1)$ and $2j$ share $W_{2j-1}$ and $R_{2j-1}$ for each $j$. This gives the required bound if $N$ is even. If $N$ is odd, let $N=2g+1$, then
\begin{align}
    \sum_{n=1}^N H(A_n^{[k]})=&H(A_1^{[k]})+H(A_2^{[k]})+H(A_3^{[k]})\notag \\
    &+\sum_{j=2}^{g} H(A^{[k]}_{2j})+H(A^{[k]}_{2j+1})\\
    \geq& H(A_1^{[k]})+H(A_2^{[k]})+H(A_3^{[k]})+(N-3)L, \label{eq:sum_path answers_odd}
\end{align}
where \eqref{eq:sum_path answers_odd} follows from \eqref{eq:pair_ans_bound} and since $2(g-1)=N-3$. If $H(A_3^{[k]})\geq L$, since $H(A_1^{[k]})+H(A_2^{[k]})\geq 2L$, we are done. Otherwise, Lemma \ref{lem:single_server_answer_entropy} applied to $S=2$, with yields
\begin{align}
    H(A_2^{[k]})\geq& \max\left\{0,2L-H(A_1^{[k]})-H(A_3^{[k]})\right\}\notag\\
    &+ 2L-H(A_3^{[k]})\label{eq:odd_path_reduction}, 
\end{align}
since $\cn(S)=\{1,3\}$. Then, \eqref{eq:odd_path_reduction} reduces to
\begin{align}\label{eq:bound_A2}
    H(A_2^{[k]})\geq 
    \begin{cases}
    2L-H(A_3^{[k]}), \,\,\,\,\, H(A_1^{[k]})+H(A_3^{[k]})\geq 2L\\
    4L-H(A_1^{[k]})-2H(A_3^{[k]}), \text{ otherwise.}
    \end{cases}
&\end{align}
Rearranging the terms in \eqref{eq:bound_A2} yields $H(A^{[k]}_1)+H(A_2^{k]})+H(A_3^{[k]})\geq 3L$ in both cases, which by substitution in \eqref{eq:sum_path answers_odd} gives the required bound for $N$ odd. This completes the proof.

\bibliographystyle{unsrt}
\bibliography{reference}
\end{document}